\documentstyle[12pt]{article}

\def\gtap{\ \raisebox{-.5ex}{\rlap{$\sim$}} \raisebox{.4ex}{$>$}\ }

\newcommand{\be}{\begin{equation}}
\newcommand{\ee}{\end{equation}}
\newcommand{\beq}{\begin{eqnarray}}
\newcommand{\eeq}{\end{eqnarray}}

\catcode`@=12
\begin{document}
\begin{titlepage}

\rightline{\vbox{\halign{&#\hfil\cr
&NTUTH-96-02\cr
&January 1996\cr}}}
\vspace{0.2in}

\begin{center}
{\large \bf  Like Sign Top Pair Production via
                $e^+e^-\to h^0A^0
                                       \to tt\bar c\bar c$
\footnote
{
Talk presented at LCWS95, 
September 8 -- 12, 1995, Morioka-Appi, Japan.
Work done in collaboration with Guey-Lin Lin.
}
}
\vfill
        {\bf George Wei-Shu HOU}\\
        {Department of Physics, National Taiwan University,}\\
        {Taipei, Taiwan 10764, R.O.C.}\\
\end{center}
\vfill
\begin{abstract}
We discuss the possibility and likelihood that one may observe
{\it like sign} top quark pair production at the Linear Collider.
In general two Higgs models, flavor changing couplings involving
top quark could be quite sizable.
Exotic neutral Higgs bosons may decay dominantly via 
$t\bar c$ or $\bar tc$ channels.
At the linear collider, $e^+e^- \to h^0A^0$ or
$H^0A^0$ production processes could lead to
$b\bar bt\bar c$, $W^+W^-t\bar c$ or
$tt\bar c\bar c$ (or $\bar t\bar tcc$)  final states.
These would mimic $T$-$\bar T$ mixing effect,
except that $T$ mesons do not even form.
\end{abstract}
\vfill
\end{titlepage}
\newpage

\section{Introduction}

In the previous workshop in Waikola, Hawaii, I presented a talk on
searching for $t\to ch^0$ decay at NLC.$^1$ 
Since then, the top quark has been found.
        With $m_t \simeq 175$ GeV, we find that BR$(t\to ch^0)$
        cannot be more than $1\%$,$^{2}$
        which is a relatively tough decay mode to study.
%
Interest in Higgs boson
induced flavor changing neutral couplings (FCNC) has also grown:
1) Lingering possibility of $m_t < M_W$;$^3$
2) Lepton number violation;$^4$
3) $CP$ violation;$^{5,6,7}$
4) $D^0$--$\bar D^0$ mixing and rare $D$ decays;$^{7,8}$
5) Generic $tcZ^0$ couplings$^9$ and effect on $bsZ$ coupling;$^{10}$
6) FCNC Higgs loop induced $e^+e^- \to \gamma^*,\ Z^* \to t\bar c$ transition;$^{11}$
7) Tree level $\mu^+\mu^- \to \mbox{neutral scalars}\to t\bar c$.$^{12}$

We are concerned here with the last two items.
Our theme is about the possibility of observing
$e^+e^- \to h^0A^0 \to tt\bar c\bar c$ or $\bar t\bar t cc$,
namely, {\it like sign top pair production at linear colliders}.
In the following, let us first see how it occurs, then contrast it
with the work of Atwood, Reina and Soni, refs. 11 and 12.

\section{The Model}

Consider the existence of two Higgs doublets,
$\Phi_1 $ and $\Phi_2$.
The ``Natural Flavor Conservation" (NFC) condition
of  Glashow and Weinberg$^{13}$ dictates that
there be just one source of mass for each fermion charge type,
usually implemented via discrete symmetries.
For example, under
$\Phi_1 \to \Phi_1$, $\Phi_2\to -\Phi_2$, one has
\begin{eqnarray}
\mbox{Model I}:\ \ \ \ \ \ \ u_R \to u_R, &&d_R\to\ \ d_R,  \nonumber \\
\mbox{Model II}:\ \ \ \ \ \   u_R \to u_R, && d_R\to -d_R.
\end{eqnarray}
In these models, called the ``Standard 2HDM(s)",
one has $\lambda_f\propto m_f$,
hence the Yukawa and mass matrices are simultaneously diagonalized,
and neutral scalar bosons are flavor diagonal {\it by construction}.
Model II is popular since it is realized in the minimal
supersymmetric standard model (MSSM).
Note also that $v_1 \equiv \langle\phi_1^0\rangle$
and $v_2 \equiv \langle\phi_2^0\rangle$
are distinct because of the discrete symetry,
hence the familiar $\tan\beta \equiv v_1/v_2$
appears in these models as a physical parameter.

Without imposing the NFC condition, i.e.,
without imposing the discrete symmetry of eq. (1),
one would have {\it two} Yukawa coupling matrices,
$\lambda^{\left(1\right)}_f$ and $\lambda^{\left(2\right)}_f$,
which in general are not proportional to
the one and only mass matrix $m_f$.
Thus,
$\lambda^{\left(1\right)}_f$ and $\lambda^{\left(2\right)}_f$
in general cannot be simultaneously diagonalized with $m_f$,
hence $\phi_1^0$ and $\phi_2^0$ would
induce FCNC at tree level.
Historically, this problem
lead Glashow and Weinberg to advocate$^{13}$ the necessity of NFC.

With spontaneous $CP$ violation,
in general$^{5}$ $\arg(v_1/v_2) \neq 0$ 
since both $v_1$ and $v_2$ are complex,
and $\tan\beta = v_1/v_2$ remains a physical parameter.
We shall, however, assume $CP$ invariance
and take both $v_1$ and $v_2$ to be real.
In this case, a linear redefinition of
$\Phi_1$ and $\Phi_2$ allows one to choose one doublet
to be the ``mass giver", which develops a
vacuum expectation value,
while the other doublet has zero vacuum expectation,
viz.$^{14}$
\begin{equation}
\langle\phi_1^0\rangle = {v\over \sqrt{2}},\ \ \ 
\langle\phi_2^0\rangle = 0.
\end{equation}
In this way, $\tan\beta$ is rotated away by the freedom
to make linear redefinitions,
and $\mbox{Re}\,\phi_1^0$ is the ``standard neutral Higgs",
except that it is {\it NOT} a mass eigenstate.
In the basis of eq. (2) and ignoring leptons,
one readily sees that
\begin{equation}
   \left(m_i^{\left(u\right)} \bar u_{iL} u_{iR}
        + m_i^{\left(d\right)} \bar d_{iL} d_{iR}\right)\;
     \left(1 + {\sqrt{2}\over v}\mbox{Re}\, \phi_1^0\right),
\end{equation}
is flavor diagonal,
however, for $\Phi_2$ Yukawa couplings, we have
\begin{eqnarray}
    && \left(\bar u_{L} \xi^{\left(u\right)} u_{R} + \bar d_{L} \xi^{\left(d\right)} d_{R}\right)\;
         \mbox{Re}\, \phi_2^0
      + \left(-\bar u_{L} \xi^{\left(u\right)} u_{R} + \bar d_{L} \xi^{\left(d\right)} d_{R}\right)\;
    \left(i\mbox{Im}\, \phi_2^0\right)  \nonumber \\
    &&  \left(-\bar d_{L} V^\dagger \xi^{\left(u\right)} u_{R} \right)\, \phi_2^-
       + \left(\bar u_{L} V\, \xi^{\left(d\right)} d_{R} \right)\, \phi_2^+ \ \ \ + H.c.
\end{eqnarray}
where $\xi^{\left(u,d\right)}$ is in general
{\it not} diagonal.
We take $V^{\left(\dagger\right)}\xi \simeq \xi$,
since the KM matrix $V \simeq 1$.

One may think$^{14}$ that the $\Phi_2$ Yukawa couplings
$\xi^{\left(u,d\right)}$ could be completely general.
However, taking cue from $V \simeq 1$,
there is a weaker form, in fact a more {\it natural} one,
for realizing ``natural flavor conservation".
Cheng and Sher observed$^{15}$ in 1987 that,
with the ansatz
\begin{equation}
\xi_{ij} \sim {\sqrt{m_i m_j}\over v},
\end{equation}
FCNC involving lower generation fermions are naturally suppressed,
without the need to push FCNC Higgs boson masses
way beyond the {\it v.e.v.} scale.
This is more natural than NFC in the following sense.
Compared to the time when Glashow and Weinberg
proposed the NFC condition,
we now know that $V\simeq 1$.
In addition,  there are two seemingly related hierarchies in nature,
namely the hierarchies in masses and KM mixing angles:
\begin{equation}
\left\{ \begin{array}{l}
m_1 \ll m_2 \ll m_3,\\
\vert V_{ub}\vert^2 \ll \vert V_{cb} \vert^2 \ll \vert V_{us}\vert^2 \ll 1,
           \end{array} \right.
\end{equation}
hence, since the KM matrix $V$ measures the ``difference"
between the $u_L$ and $d_L$ diagonalisation matrices,
one expects from naturalness that 
\begin{equation}
\xi_{ij} = {\cal O}(V_{i3}V_{j3})\, {m_3\over v},
\end{equation}
unless fine-tuned cancellations are implemented.
It was from this perspective that we exphasized$^2$
the pertinence and importance of FCNC Higgs induced transitions
involving the heaviest quark, the top.
Before we turn to low energy constraints,
some further formalism is necessary.

So far we have been in the ``weak" basis. We need to work in
the (scalar) mass basis, which is determined by the
the Higgs potential $V(\Phi_1,\ \Phi_2)$.
In fact, we need not care about the details of $V(\Phi_1,\ \Phi_2)$,
since electric charge and $CP$ invariance dictates that
only $\mbox{Re}\, \phi_1^0$ and $\mbox{Re}\, \phi_2^0$
can mix, that is
\begin{equation}
  \begin{array}{ccc}
    \mbox{\underline{Gauge basis}} & & \mbox{\underline{Mass basis}} \\
                                   \\
    \mbox{Re}\, \phi_1^0,\ \mbox{Re}\, \phi_2^0 &
                   \ \ \ \ \ \ \ {\Longrightarrow}\ \ \ \ \ \ \ & H^0,\ h^0 \\
    \mbox{Im}\, \phi_2^0 & 
                                               {\longrightarrow} & A^0 \\
                         \phi_2^\pm & 
                                               {\longrightarrow} & H^\pm. 
  \end{array}
\end{equation}
The effect of the Higgs potential $V(\Phi_1,\ \Phi_2)$
can be summarized in the rotation
\begin{equation}
 \left( \begin{array}{l}
             H^0 \\  h^0
          \end{array} \right)
  =  \left( \begin{array}{rr}
                  \cos\alpha & \sin\alpha  \\ 
                 - \sin\alpha & \cos\alpha 
               \end{array} \right)
 \left( \begin{array}{l}
             \sqrt{2}\, \mbox{Re}\, \phi_1^0 \\  \sqrt{2}\, \mbox{Re}\, \phi_2^0
          \end{array} \right),
\end{equation}
where the neutral scalar rotation angle $\sin\alpha$
is a physical parameter of the model.
In the limit that $\sin\alpha \to 0$, 
one has
\begin{equation}
\left\{ \begin{array}{l}
H^0 \ \ \leadsto \ \ \sqrt{2}\, \mbox{Re}\, \phi_1^0\\
h^0\; \ \ \leadsto \ \ \sqrt{2}\, \mbox{Re}\, \phi_2^0
           \end{array} \right. \ \ \ \ \ \ \ \ \ \ \ \ \ \ (\sin\alpha \longrightarrow 0),
\end{equation}
where $H^0$ is now the ``standard" Higgs boson with
flavor diagonal couplings,
while $h^0$ in this limit has general Yukawa couplings
(subject to eqs. (5) or (7)) but does not couple to vector bosons
or charged Higgs bososn.
We note that our convention here for $H^0$ and $h^0$
differs from the usual convention in MSSM.

In the following, we shall take the scenario of eqs. (5) or (7) for
Yukawa couplings, calling it Model III,
and concentrate on consequences of
the limiting case of eq. (10), which is also the simplifying
assumption taken in refs. 11, 12 and 14. 
In the end, the consequences of $\sin\alpha \neq 0$ will also
be discussed.


\section{Low Energy Constraints}

Low energy FCNC processes involve external quarks
belonging to lower generations.
One readily sees from eqs. (5) or (7) that constraints on 
Higgs masses in Model III are far less stringent than in Models I and II.
Specifically, important constraints come from
$d$ type quark sector and charged leptons.

$K$--$\bar K$ and $B$--$\bar B$ mixings
were considered by Cheng and Sher$^{15}$
and Sher and Yuan$^{16}$, 
assuming eq. (5), leading to a
not so stringent bound of $m_{h^0 } \gtap 80$ GeV
for $h^0$, and a somewhat more stringent bound on $A^0$.
Note, however, that the mass bound would weaken
if $\xi$ is weaker than that given by eq. (5).
%


As originally noted by Bjorken and Weinberg$^{18}$,
because of the need to have 3 chirality flips, the 
$h^0$ or $A^0$ induced one-loop contribution to $\mu\to e \gamma$
is rather suppressed. 
At two-loop order, then,
the one-loop effective scalar--$\gamma$--$\gamma$ coupling
induces $\mu\to e \gamma$ transition with just 1 chirality flip,
allowing this process to dominate over the one-loop process.
For $m_t \simeq 175$ GeV, we find$^{17}$
that 
$m_{h^0,\; A^0} \gtap 150\ \mbox{GeV}$,
i.e. of order $v$ scale, which is quite reasonable,
and roughly agrees with the
$K$--$\bar K$ and $B$--$\bar B$ mixing bounds.


Within Model II and to leading-log 
order in QCD corrections, CLEO finds$^{19}$
$m_{H^+} \gtap 250$ GeV.
This lower bound is specific to Model II
because of a $\tan\beta$ independent contribution.$^{20}$.
In Model III,
bounds could weaken because of the remaining freedom 
in $\xi^{\left(u\right)}$ and $\xi^{\left(d\right)}$.
Furthermore, 
it has been argued$^{21}$ that inclusion of next-to-leading order
QCD corrections tends to soften the bound.
Thus, we take $m_{H^+} \gtap 150 - 250$ GeV
as  a reasonable lower bound.
This bound is rather consistent with the bound
on FCNC neutral scalar bosons.


The upshot is, it  is rather likely that
\begin{equation}
v\sim m(\mbox{FCNC Higgs}) \gtap m_t.
\end{equation}

\section{Decay Scenario and Production Processes}

There is no $A^0VV$ coupling to start with, where
$V = W$ or $Z$.
In the limit of eq. (10), i.e. $\sin\alpha \rightarrow 0$,
there is also no $h^0VV$ coupling.
Taking at face value the lower bound of eq. (11),
we restrict ourselves to the kinematic domain of
\begin{equation}
200\ \mbox{GeV} < m_{h^0,\; A^0} < 2m_t \simeq 350\ \mbox{GeV},
\end{equation}
then $h^0$ and $A^0$ can only decay via
$t\bar c$ (or $\bar tc$) and $b\bar b$.
We note that $2m_t \simeq 350$ GeV is roughly 
the Higgs boson mass reach for a 500 GeV Linear Collider (NLC).
The lower range of 200 GeV is chosen such that
$h^0,\ A^0 \to t\bar c$ decay is not overly restricted by phase space.
That is, 
\begin{equation}
    {\Gamma(h^0,\ A^0 \to t\bar c + \bar tc)\over 
                         \Gamma(h^0,\ A^0 \to b\bar b)}
       \cong  {2\xi^2_{ct}\over \xi^2_{b\bar b}}\, 
                \left[1 - {m_t^2\over m_{h^0,\, A^0}^2}\right]^2
       \gtap  2{m_cm_t\over m_b^2}\,
                \left[1 - \left({175\over 200}\right)^2\right]^2 \gtap 1.5.
\end{equation}
Because of the large top quark mass,
the first factor is of order 25 and large. 
The second phase space factor increases rapidly for $m_{h^0,\, A^0} > 200$ GeV.
We therefore conclude that,
for $m_{h^0,\, A^0} \in (200,\ 350)$ GeV,
which is a very reasonable domain,
$h^0,\ A^0 \to t\bar c + \bar tc$ could likely be dominant
over the $b\bar b$ mode.

The production processes are rather standard.
Again, in the limit of eq. (10) with $\sin\alpha \rightarrow 0$,
the only neutral Higgs boson that couples to vector boson
pairs is the ``standard", flavor diagonal $H^0$ boson.
Thus,
the processes
\begin{eqnarray}
e^+e^- & \rightarrow & Z^* \rightarrow H^0Z^0, \\
e^+e^- & \rightarrow & \nu\bar\nu + H^0 \ \ \ \ \ \ \  (WW\ \mbox{fusion}),
\end{eqnarray}
are both $\propto \cos\alpha$ in amplitude,
and would appear to be completely standard.
The nonstandard $h^0$ boson would not be produced
since it is $\propto \sin\alpha$ in amplitude.
However, as is well known, the associated production process
\begin{equation}
e^+e^- \ \ \rightarrow \ \ Z^* \ \ \rightarrow \ \ h^0A^0
\end{equation}
is also $\propto \cos\alpha$ in
amplitude, and has a cross section similar to 
the process of eq. (14), except for difference in phase space
and the transverse $Z$ contribution proportional to $M_Z^2/s$.
Since $h^0,\ A^0 \to t\bar c + \bar tc$  is dominant
over $b\bar b$, and because of the real field nature of
$h^0$ and $A^0$ (and $CP$ conservation),
they each decay to $t\bar c$ and $\bar tc$ with
equal weight.
Thus, for the process of eq. (16), one could find
$50\%$ of the cross section going into 
{\it like sign top pair} events,
namely,
\begin{equation}
\sigma(e^+e^- \ \rightarrow \ tt\bar c\bar c + \bar t\bar tcc)
\sim 0.5\times \sigma(e^+e^- \ \rightarrow \ h^0A^0).
\end{equation}

There is one special process that deserves mentioning.
It is now accepted that having back-scattered laser beams
to convert linear $e^+e^-$ colliders into
effective $\gamma\gamma$ colliders of almost equal energy
is something highly desirable$^{22}$ when the NLC is built.
The chief reason is the interest in
{\it measuring} the $H\to \gamma\gamma$ width,
because of its sensitivity to beyond the standard model effects.$^{23}$
In the case of eq. (10), 
only the top contributes, with
\begin{equation}
\sigma(\gamma\gamma \rightarrow h^0,\ A^0) \sim 1\ \mbox{fb},
\end{equation}
which provides for clean FCNC single top production.


\section{Rough Numbers and Background}

For  the mass range of eq. (12) with $m_{A^0} > m_{h^0}$,
we list in Table 1 the number of events  ${\cal N}(h^0 A^0)$ for a
500 GeV linear collider with $\int {\cal L}\, dt = 50$ fb$^{-1}$.
\vskip 0.4cm
\begin{center}
{Table 1: Number of events in $h^0A^0$ channel for 500 GeV NLC at 50 fb$^{-1}$.}
\vskip 0.4cm
\begin{tabular}{|c|r|r|r|r|r|r|r|r|} \hline
  $m_{A^0}$ (GeV)  & 200 & 250 & 300 & 200 & 250 & 300
                                                                                    & 200 & 250     \\ \hline
  $m_{h^0}$ (GeV)  & 100 & 100 & 100 & 150 & 150 & 150
                                                                                    & 200 & 200     \\ \hline
  ${\cal N}(h^0 A^0)$            & 1100 & 750 & 600 & 900 & 500 & 150
                                                                                    & 500 & 200     \\ \hline
\end{tabular}
\end{center}
\vskip 0.2cm
It is clear that, when phase space permits, one expects
the raw number of events at the $10^3$ order.
In contrast, $H^0 Z^0$ associated production results in 
$\sim 3\times 10^3$ -- $500 $ events for 
the mass range $m_{H^0} \in (150,\ 300)$ GeV.
We list the number of 
potential background events in Table 2.
\vskip 0.4cm
\begin{center}
{Table 2: Number of raw events in potential background modes.}
\vskip 0.4cm
\begin{tabular}{|c|c|c|} \hline
  \ \ ${\cal N}(t\bar t)$\ \     &   ${\cal N}(W^+W^-)$    &   ${\cal N}(Z^0Z^0)$     \\ \hline
         $\sim 30,000$           &        $\sim 400,000$        &       $\sim 30,000$         \\ \hline 
\end{tabular}
\end{center}
\vskip 0.2cm
From the last two columns of Table 1, taking ${\cal N}(h^0A^0)$ to be of order
500, one expects of order 250 $tt\bar c\bar c$ or $\bar t\bar tcc$ events,
resulting in $\sim$ 12 signal events in the 
\begin{equation}
e^+e^- \longrightarrow \ell^\pm\ell^{\prime\pm} + \nu\nu + 4j
\end{equation}
channel, where the 4 jets have flavor $bb\bar c\bar c$ or $\bar b\bar bcc$. 
With a good detector, in part thanks to the large top quark mass,
this distinctive signature has seemingly {\it no background}.
In contrast, single $\ell + \nu + 6j$ events 
or opposite sign dilepton events from $t\bar t c\bar c$ final states
would be swamped by background listed in Table 2,
which are orders of magnitude higher.
In particular,
standard $e^+e^- \to t\bar t$ pair production
with additional gluon radiation may be especially irremovable.


\section{Discussion}

Although like sign top pair production is quite intriguing,
the limiting case of eq. (10) may be a little extreme.
Loosening the condition so $\sin\alpha \neq 0$ results in
$H^0$--$h^0$ mixing.
Both $H^0Z^0$ and $h^0Z^0$ associated production
become possible, with the respective weights
of $\cos^2\alpha$ and $\sin^2\alpha$.
Assuming that $\sin^2\alpha < \cos^2\alpha$,
we note that $H^0$ width is in general still large,
since it is dominated by $H^0 \to W^+W^-$ and $ZZ$,
and $H^0\to t\bar c + \bar tc$ would be rather rare.
For the $e^+e^- \to h^0Z^0$ porcess, the cross section is
suppressed by $\sin^2\alpha$, but, when $\sin^2\alpha$ grows,
the $h^0\to t\bar c$ mode is quickly overwhelmed by the
$h^0\to VV$ modes, and again the effective cross section is
reduced. In any case, one expects some events in
$Z + t\bar c$ production, with large fraction into $b\bar b t\bar c$.

We have also listed in Table 1 the possibility of $m_{A^0} > m_t$ but
$m_{h^0} < m_t$. If such is the case, we expect
$h^0\to b\bar b$, and again we have
$e^+ e^- \to b\bar b t\bar c$, with a different $b\bar b$ pair mass.
With good $b$-tagging efficiency, these modes should not be difficult to study.

We now compare with the results of Atwood, Reina and Soni.
For $e^+e^- \to \gamma^*,\; Z^* \to t\bar c$ via
$h^0,\ A^0$ loop effects,
Atwood et al.$^{11}$ find less than $0.1$ event for a 500 GeV NLC
with $\int{\cal L}\, dt = 50$ fb$^{-1}$.
For heavy $h^0$ and $A^0$ where
our processes are kinematically forbidden,
the loop induced cross section also 
goes down by another order of magnitude.$^{11}$
Thus, this process is unlikely to be observable at the NLC.
Loop suppression in this case is no match against
normal tree level processes that we discuss.

For $\mu^+\mu^- \to h^0$, $A^0 \to t\bar c + \bar tc$,
the process occurs at tree level and has a sizable cross section.$^{12}$
But in the scenario (of eq. (10)) taken, 
because of the absence of $h^0,\; A^0 \to VV$ decay mode,
one needs a rather fine stepped energy scan because of the
narrowness of the $h^0$ and $A^0$ width.
Together with the uncertainty of whether a high energy,
high luminosity $\mu^+\mu^-$ collider can be realized,
we feel that this process is less straightforward to study
than the processes we discuss for the NLC.
In particular, it may be even less promising than
searching for $\gamma\gamma \to h^0,\ A^0$ at the NLC. 

The signature of like sign top pair production
is rather analogous to producing $B^0B^0$ or $\bar B^0\bar B^0$
final states via $B^0\bar B^0$ pair production followed
by $B^0$--$\bar B^0$ mixing.
However, for $m_t \simeq 175$ GeV,
top mesons (be it $T_u^0$ or $T_c^0$) do not even form!
The reason that we get like sign top pair production here
is due to the real neutral scalar field nature of $h^0$ and $A^0$,
which circumvents the usual condition of associated production
of $t\bar t$ pairs in most processes.
Since $h^0$ and $A^0$ contribute to $B$--$\bar B$ mixings,
the like sign top pair production effect reported here
is related to neutral meson--anti-meson mixing phenomena.
We know of no other way to make $tt$
or $\bar t\bar t$ pairs in an $e^+e^-$ collider environment.


\section{Summary}

In the context of a two Higgs doublet model without imposing NFC condition,
neutral scalar bosons in general has FCNC couplings.
We have presented the case where
$h^0$, $A^0 \rightarrow t\bar c + \bar tc$ could be the
dominant decay mode, 
for the rather reasonable mass range
$m_{h^0,\, A^0} \in (200,\ 350)$ GeV.
The most intriguing consequence is the possibility of 
detecting like sign top pair production
via $e^+e^-\to h^0A^0 \to tt\bar c\bar c$ or $\bar t\bar tcc$.
One may also detect single top FCNC production 
via $\gamma \gamma \to h^0,\ A^0 \to t\bar c + \bar tc$.
The situation becomes richer if $\sin\alpha \neq 0$,
where one has $h^0$--$H^0$ mixing,
leading to $Z + t\bar c$ events.
The situation could get even richer if
$CP$ is violated in Higgs sector, where one
has $h^0$--$H^0$--$A^0$ mixing.
Since the Higgs sector of minimal SUSY (MSSM) is flavor diagonal, 
{\it observation of the signatures reported here
would rule out MSSM}.
We urge experimental colleagues to 
study the signal vs. background issue carefully.


\section{References}

\begin{thebibliography}{9}
\bibitem{1} G.W.S. Hou, Proceedings of  
{\it Workshop on Physics and Experiments with Linear $e^+e^-$ Colliders},
eds. F.A. Harris et al. (World Scientific, 1993).
%
\bibitem{2} W.S. Hou, {\it Phys.~Lett.} {\bf B296} (1992) 179. 
%
\bibitem{3} W.S. Hou, {\it Phys.~Rev.~Lett.} {\bf 72} (1994) 3945.
%
\bibitem{4} T.S. Kosmas, G.K. Leontaris and  J.D. Vergados,
 {\it Prog.~Nucl.~Part.~Phys.} {\bf 33} (1994) 397;
L. Wolfenstein and Y.L. Wu, CMU-HEP-94-26, 1994.
%
\bibitem{5} L. Wolfenstein and Y.L. Wu, 
{\it Phys. Rev. Lett.} {\bf 73} (1994) 1762 and 2809.
%
\bibitem{6} M. Masip and A. Rasin, UMD-PP-95-143 and UMD-PP-96-17.
%
\bibitem{7} G. Blaylock, A. Seiden and Y. Nir, 
{\it Phys. Lett.} {\bf B355} (1995) 555;
L. Wolfenstein,  {\it Phys. Rev. Lett.} {\bf 75} (1995) 2460.
%
\bibitem{8} J.L. Hewett, talk given at Workshop on
{\it Tau Charm Factory in the era of B Factories and CESR}, 
Stanford, August 1994;
Y. Nir, talk given at
{\it 6th International Symposium on Heavy Flavor Physics},
Pisa, June 1995. 
%
\bibitem{9} T. Han, R.D. Peccei and X. Zhang, {\it Nucl.~Phys.} {\bf B454} (1995) 527.
%
\bibitem{10} E. Nardi, WIS-95-40-PH, August 1995. 
%
\bibitem{11} D. Atwood, L. Reina and A. Soni, SLAC-PUB-95-6927, June 1995.
%
\bibitem{12} D. Atwood, L. Reina and A. Soni, {\it Phys. Rev. Lett.} {\bf 75} (1995) 3800.
%
\bibitem{13} S.L. Glashow and S. Weinberg, 
{\it Phys. Rev.} {\bf D15} (1977) 1958.
%
\bibitem{14} M. Luke and M.J. Savage,
{\it Phys. Lett.} {\bf B307} (1993) 387.
%
\bibitem{15} T.P. Cheng and M. Sher,
{\it Phys. Rev.} {\bf D35} (1987) 3484.
%
\bibitem{16} M. Sher and Y. Yuan,
{\it Phys. Rev.} {\bf D44} (1991) 1461.
%
\bibitem{17} D. Chang, W.S. Hou and W.-Y. Keung,
{\it Phys. Rev.} {\bf D48} (1993) 217.
%
\bibitem{18} J. D. Bjorken and S. Weinberg,
{\it Phys. Rev. Lett.} {\bf 38} (1977) 622.
%
\bibitem{19} M.S. Alam et al. (CLEO Collab.),
{\it Phys.~Rev.~Lett.} {\bf 74} (1995) 2885.
%
\bibitem{20} B. Grinstein and M.B. Wise,
{\it Phys. Lett.} {\bf B201} (1988) 274;
W.S. Hou and R.S. Willey,
{\it ibid.} {\bf B202} (1988) 591.
%
\bibitem{21} M. Ciuchini, talk given at  27th International Conference
on High Energy Physics, Glasgow, July 1994.
%
\bibitem{22} D. Miller, plenary talk on ``Other Options", this Proceedings.
%
\bibitem{23} I. Watanabe, talk in ``Other Options" parallel session, this Proceedings.

\end{thebibliography}
\end{document}